\documentclass[aps,twocolumn,noshowkeys,nofootinbib]{revtex4-2}
\usepackage{amsmath}
\usepackage{graphicx}
\usepackage{bm}
\usepackage{hyperref}
\usepackage{xcolor}
\usepackage{multirow}

\newcommand{\Eref}[1]{Eq.~(\ref{#1})}

\begin{document}

\title{Relativistic Fock space coupled cluster study of bismuth electronic structure to extract the Bi nuclear quadrupole moment}

\date{08 June 2021}

\begin{abstract}
We report the value of the electric quadrupole moment of $^{209}$Bi extracted from the atomic data. For this, we performed electronic structure calculations for the ground $^4S^o_{3/2}$ and excited $^2P^o_{3/2}$ states of atomic Bi using the Dirac-Coulomb-Breit Hamiltonian and the Fock space coupled cluster method with single, double, and full triple amplitudes for the three-particle Fock space sector. The value of the quadrupole moment of $^{209}$Bi, $Q(^{209}$Bi$)=-418(6)$~mb, derived from the resulting electric field gradient values and available atomic hyperfine splittings is in excellent agreement with molecular data. Due to the availability of the hyperfine constants for unstable isotopes of Bi, current atomic calculation allows also to correct their quadrupole moments.
\end{abstract}

\author{L.V.\ Skripnikov$^{1,2}$}
\email{skripnikov\_lv@pnpi.nrcki.ru,\\ leonidos239@gmail.com}
\author{A.V. Oleynichenko$^{1,3}$}
\author{A.V. Zaitsevskii$^{1,3}$}
\author{D.E. Maison$^{1,2}$}
\author{A.E.\ Barzakh$^{1}$}
\affiliation{$^{1}$Petersburg Nuclear Physics Institute named by B.P.\ Konstantinov of National Research Center ``Kurchatov Institute'' (NRC ``Kurchatov Institute'' - PNPI), 1 Orlova roscha, Gatchina, 188300 Leningrad region, Russia}
\homepage{http://www.qchem.pnpi.spb.ru}
\affiliation{$^{2}$Saint Petersburg State University, 7/9 Universitetskaya nab., St. Petersburg, 199034 Russia}
\affiliation{$^{3}$Department of Chemistry, M.V. Lomonosov Moscow State University, Leninskie gory 1/3, Moscow, 119991~Russia}

\maketitle

\section{Introduction}
The quadrupole moment $Q$ is one of the main properties of the nucleus with an intrinsic spin $I\ge1$
\footnote{In this paper we consider the spectroscopic quadrupole moment. Note that one should discriminate it from the intrinsic quadrupole moment $Q_0$ which is not equal to zero even for $I=1/2$ nuclei.}. It characterizes the non-sphericity of the nuclear charge distribution.
Nuclei with quadrupole and octupole deformations are good candidates to measure the nuclear Schiff moment, which violates time-reversal and spatial parity symmetries of fundamental interactions and can be used to search for the new physics~\cite{Auerbach:1996,Spevak:97,Skripnikov:2020c}. In deformed nuclei, the Schiff moment can be strongly enhanced.

The prevailing methods to obtain $Q$ are atomic and molecular spectroscopies. The interaction of the nuclear quadrupole moment (NQM) with the electric field gradient (EFG) at the nucleus leads to the shift of the hyperfine structure components in atomic and molecular spectra. The electric quadrupole hyperfine structure parameter $B$, which can be extracted from the experimental data, is proportional to the NQM value. Therefore, an accurate theoretical prediction of the EFG value enables one to extract the value of $Q$ using the following relation:
\begin{equation}
Q[{\rm b}]=\frac{B {\rm [MHz]}}{234.9648867~q[{\rm a.u.}]},
\end{equation}
where $q$ is the electric field gradient in a.u.~($=E_H/a_B^2$); $E_H=1$~hartree and $a_B$ is the Bohr radius. In many cases, very accurate data for atomic $B$ constants are available. However, often the accuracy of the extracted NQM values is limited by the accuracy of the theoretical prediction of EFG.

The highly charged ions with the $^{209}$Bi nucleus have been recently used to test predictions of quantum electrodynamics which resulted in the so-called (magnetic dipole) ``hyperfine puzzle'' due to the disagreement between the theoretical predictions~\cite{Shabaev:01a,Volotka:12} and experiment~\cite{Ullmann:17}. More recently, it has been shown~\cite{Skripnikov:18a} that the discrepancy was caused by the inaccurate tabulated value of the magnetic dipole moment of $^{209}$Bi. The $^{209}$Bi nucleus has the spin $I=9/2$ and also possesses the NQM. Previously, a large set of the values of $Q$ in the range from -370 mb to -710 mb, has been obtained, see Ref.~\cite{BieronPRL:01} for a review. The advanced, up to date atomic calculations gives $Q=-516(15)$~mb~\cite{BieronPRL:01}, whereas in the recent molecular studies values lower by 20\% were obtained: $Q=-420(8)$~mb~\cite{Quevedo:2013} and $Q=-415.1$~mb~\cite{Avijit:2016}. Such a large discrepancy demands a reconciliation since the experimental uncertainties are in many cases less than 5\%. We believe that the main sources of this discrepancy are the drawbacks of atomic calculations, as the neutral bismuth atom has three unpaired electrons and strong static and dynamic electron correlation effects. Therefore, the accurate prediction of EFG in the atomic Bi case is especially a challenging problem.

In the present paper, we apply the Fock space relativistic coupled cluster theory with single, double, and full iterative triple cluster amplitudes developed in our group to extract the value of EFG for the $^{209}$Bi nucleus from the atomic data. Moreover, we recalculate the $Q$ values for a number of radioactive isotopes where corresponding atomic data are available~\cite{Schmidt:2018,Pearson:2000}.

\section{Fock space coupled cluster theory for the three-particle sector}

In the present paper, we focus on the ground $^4$S$^o_{3/2}$ and excited $^2$P$^o_{3/2}$ states of the Bi atom. For both states with the same relativistic symmetry, several configurations $(6p^3)_{3/2}$ contribute to wavefunctions with comparable weights and thus should be treated on equal footing. Therefore the use of inherently multireference approaches to describe these states seems \emph{a priori} advantageous. 

An accurate treatment of electronic structure of heavy-element atoms and molecules requires to describe simultaneously both the relativistic and electronic correlation effects. For the electronic states with two or more open-shell electrons the static correlation and its interplay with the dynamic correlation effects has to be thoroughly analyzed and accounted for. One of the most efficient and reliable methods allowing one to treat all these effects in a balanced manner is based on the relativistic multireference coupled cluster theory in its Fock space formulation (FS-RCC) \cite{Kaldor:1991,Bartlett:2007}. The most important feature of FS-RCC consists in the fact that all the reference determinants are treated on equal footing allowing one to properly capture the most important static correlation effects. Up to now, the applications of FS-RCC were limited to the systems with no more than two ``valence’’ electrons (unpaired electrons over the Fermi vacuum state) \cite{EliavHess:1998,Visscher:2001}. The attempts to extend the scope of applicability of its non-relativistic counterpart to systems with three or even more (up to six) electrons were made by Hughes and Kaldor in early 1990s  \cite{Haque:1985,Hughes:1992,Hughes:1993,Hughes:1993-review,Hughes:1995} and recently by Meissner and co-authors \cite{Meissner:2020}. Though the reported pilot applications were restricted to atoms and simple molecules of second- and third-row elements, the results have clearly shown that the inclusion of contributions of triple excitation operators is inevitable to achieve a highly accurate modelling.

The Fock space coupled cluster theory implies the use of a series of model subspaces (Fock space sectors) obtained by distributing various numbers of valence electrons (in our case from 0 to 3) among ``active’’ spinors. These subspaces form the model space. The many-electron wavefunctions $\psi_i$ are approximated by
\begin{equation}
    \psi_i = \{ \exp(T) \}_N \tilde{\psi_i},
    \label{exponential}
\end{equation}
where the model vectors $\tilde{\psi}$ are projections of the corresponding wavefunctions onto the model space, $\{ \}_N$ denotes the normal ordering with respect to the chosen Fermi vacuum and the cluster operator $T$ is given by a linear combination of excitation operators. These operators are classified according their rank $r$, and the numbers of destructed active particles $n_{p} $
\begin{equation}
T=\sum_{ n_{p}, r } \, T^{(n_{p})}_r.
\label{sectorrank}
\end{equation}
The sum (\ref{sectorrank}) is truncated at some maximum $r$ value (2, model with singles and doubles, FS-CCSD, or 3, including additionally triples, FS-CCSDT). In spite of the truncation, the exponential form of Eq.~(\ref{exponential}) ensures the incorporation of all higher excitations into the wavefunction. This feature is known to be of high importance for proper treatment of dynamic correlation effects~\cite{Bartlett:2007}. Furthermore, the use of Ansatz (\ref{exponential}) is essential for the exact size-consistency of the results obtained with truncated cluster operators; the latter feature is essential for the present problem which requires to correlate all 83 electrons of neutral Bi.

The equations for the cluster amplitudes can be found, for example, in Refs.
\cite{Kaldor:1991,Bartlett:2007}. Once the amplitudes are determined, the effective Hamiltonian $\tilde{H}$ matrix is constructed and diagonalized to obtain energies of electronic states as eigenvalues and model vectors $\tilde{\psi}$ as eigenvectors:
\begin{equation}
\tilde{H}=\left( \overline{H \,\{ \exp(T) \}_N}\right) _{Cl},\label{effham}%
\end{equation}
where $H$ is the many-electron Hamiltonian (e.g. the relativistic Dirac-Coulomb-Breit one), the subscript $Cl$ marks the closed part of an operator and the overbar denotes its connected part.

Amplitude equations are solved subsequently for the Fock space sectors, starting from the zero active particle (vacuum) sectors. Within the CCSD approximation, all amplitudes are fully defined by the equations for the vacuum, one-particle and two-particle sectors; nevertheless, the effective Hamiltonian (\ref{effham}) for the three-particle sector can be constructed using the single and double excitations completely inherited from the lower sectors \cite{Hughes:1992,Hughes:1993-review}. Thus the FS-CCSD model suffers from the rather rough accounting for differential correlation associated with the addition of a third active electron: this effect is partially included \emph{via} ``disconnected''  products of single and double excitations. The lowest-rank excitations with amplitudes which are specific for the system with three active particle are triple excitations destroying all three valence particles; therefore one can suppose that the incorporation of triples is even more important than for the lower sectors where the non-perturbative account of triple excitations (the FS-CCSDT model) have been demonstrated indispensable for acheiving the meV-level of accuracy for excitation energies \cite{Oleynichenko:20}. Unfortunately, the full FS-CCSDT model is too computationally demanding for most practical applications. The accuracy of the FS-CCSD model is believed to be sufficient for capturing the bulk of dynamic correlation effects and then the correction on triples can be introduced by means of the additive scheme based on the FS-CCSDT calculations in a reduced basis for a smaller number of correlated electrons (similar to that employed in \cite{Oleynichenko:20}). 

The relativistic FS-CCSD and FS-CCSDT models for three open-shell electrons were implemented in the {\sc exp-t} program package \cite{Oleynichenko_EXPT,EXPT_website}. The details of this implementation will be published elsewhere.

\section{Computational details}
In correlation calculations we have used the Dirac-Coulomb and Dirac-Coulomb-Breit Hamiltonians. They are given by the following expression:
\begin{equation}
    H_{\textrm{el}} = 
     \Lambda^{(+)}
     \left[
    \sum\limits_{i=1}^{N_e} h(i) +   
    \mathop{{\sum}}_{i<j}^{N_e} 
        V_{ij}
    \right]
\Lambda^{(+)},
\label{Ham}
\end{equation}
where $h(i)$ is the one-particle Dirac Hamiltonian and $V_{ij}$ is the interelectronic interaction. Indices $i$ and $j$  run over all electrons of the system, $r_{ij}$ is the distance between $i$-th and $j$-th electrons, $\boldsymbol{\alpha}_i$ is the vector of Dirac $\alpha$-matrices, $\Lambda^{(+)}$ are projectors on the positive-energy states. In the Dirac-Coulomb case $V_{ij}=1/r_{ij}$. In the frequency-independent Breit approximation $V_{ij}$ is given by the following expression:
\begin{equation}
\label{Hb}
    V_{ij}
    = \frac{1}{r_{ij}}
    - \frac{
                (\bm{\alpha}_i \cdot \bm{\alpha}_j)
            }{r_{ij}}
    -\frac{(\bm{\alpha}_i \cdot \bm{\nabla}_{i})(\bm{\alpha}_j \cdot \bm{\nabla}_{j}) r_{ij}}{2},
\end{equation}
where $\bm{\nabla}_i = \left( \frac{\partial}{\partial x_i}, \frac{\partial}{\partial y_i}, \frac{\partial}{\partial z_i} \right)$ is the gradient operator over $i$-th electron coordinates. The second term in \Eref{Hb} stands for the magnetic interaction of electrons and is known as the Gaunt term. The last terms represents the retardation correction to the instantaneous Coulomb potential. 

The main contribution to EFG has been obtained within the relativistic Fock space coupled cluster method with single and double cluster amplitudes using the Dirac-Coulomb Hamiltonian and finite-field approach~\cite{Monkhorst:77} for property calculation. The target $6s^26p^3$-states of the neutral Bi atom were achieved in the $0h3p$ Fock space sector; the closed-shell $6s^2$ Bi${}^{3+}$ cation was considered as the Fermi vacuum state (the $0h0p$ sector) and the $6p$-spinors comprised the active space. Note that for this minimal active space the intruder state problem does not arise and no additional special techniques to suppress it are required. For this calculation we have used basis set that contains 43$s-$, 37$p-$, 25$d-$, 17$f-$, 10$g-$, 8$h-$ and 6$i-$ Gaussian functions. This basis set has been obtained from the uncontracted Dyall's AAEQZ basis set~\cite{Dyall:12} for Bi by augmenting it with tight and diffuse $s$-,$p$-,$d$-,$f$-,$g$-functions and addition of $h-$ and $i-$ type functions using the procedure developed by us in Refs.~\cite{Skripnikov:2020e,Skripnikov:13a}. Correlation effects for all 83 electrons have been considered. In the correlation calculations we have included all virtual orbitals with one-particle energies up to 10000 Hartree which is important for core properties~\cite{Skripnikov:17a,Skripnikov:15a}. Table~\ref{TRes} also gives estimated contribution of the basis set components with $l>6$ missing in the FS-CCSD calculations. This correction has been obtained using the extrapolation procedure which takes into account contribution of harmonics with $l=5,6$. 

To take into account the contribution of triple cluster amplitudes, the FS-CCSDT method in the $0h3p$ sector was employed. In order to make calculations feasible the basis set consisting of 36$s-$, 30$p-$, 18$d-$, 13$f-$, 6$g-$, 2$h-$ and 1$i-$type functions was used. This basis set corresponds to the Dyall's AAETZ basis set~\cite{Dyall:06} augmented with tight and diffuse $s-$, $p-$functions and natural $g-$, $h-$ and $i-$ type functions~\cite{Skripnikov:2020e,Skripnikov:13a}. In this calculations 60 inner-core electrons were excluded from correlation treatment. According to our tests, the contribution of triple cluster amplitudes is rather stable with respect to variation of the basis set size and active space size. About $4\cdot 10^8$ unique cluster amplitudes standing in the exponent (\ref{exponential}) were optimized, suggesting good level of accounting for dynamic electron correlation effects.

The Breit interaction contribution has been calculated at the FS-CCSDT level with excluded 60 inner-core electrons as in the calculation described above and the basis set consisting of 36$s$-, 30$p$-, 18$d$-, 13$f$-, 4$g$-, 1$h$-type functions based in the Dyall's AAETZ basis set~\cite{Dyall:06}. 

Dirac(-Gaunt)-Hartree-Fock calculations of the Bi bispinors were carried out using the  {\sc dirac15} software \cite{DIRAC15} for the neutral Bi state. All coupled cluster calculations were performed within the {\sc exp-t} code \cite{Oleynichenko_EXPT, Oleynichenko:20}. Calculations of matrix elements of the Breit interaction and four-index transformation of these matrix elements have been performed within the code developed in Ref.~\cite{Maison:2019,Maison:20a}.

\section{Results and discussion}

\begin{table}[]
\caption{The calculated values of the electric field gradient in a.u.($={\rm E_H/a_B^2}$) for the ground $^4$S$^o_{3/2}$ and excited $^2$P$^o_{3/2}$ electronic states of neutral bismuth and the deduced values of the NQM of $^{209}$Bi.}
\begin{tabular}{lrr}
\hline
\hline
                              & $6p^3$ $^4$S$^o_{3/2}$ & $6p^3$ $^2$P$^o_{3/2}$ \\
\hline                              
                              & \multicolumn{2}{c}{EFG:}       \\
FS-CCSD                       & 2.983       & -10.292          \\
basis set correction          & 0.055       & -0.050           \\
FS-CCSDT$-$FS-CCSD            & 0.117       & 0.276            \\
Breit contribution            & -0.058      & 0.088            \\
Total                         & 3.097       & -9.978           \\
                              &             &                  \\
$B$, MHz~\cite{HullBrink:1970} & -305.067(2) & 978.638(10)    \\
$Q(^{209}$Bi), mb             & -419        & -417             \\             
\hline
\hline
\end{tabular}
\label{TRes}
\end{table}

Electric quadrupole hyperfine constants are known for different electronic states of Bi. In the present paper we consider the ground $6p^3$ $^4$S$^o_{3/2}$ and the excited $6p^3$ $^2$P$^o_{3/2}$ electronic states as the most precise experimental data for the hyperfine constants $B$ of $^{209}$Bi are available for them~\cite{HullBrink:1970}. 

The calculated values of EFG for these two states are given in Table~\ref{TRes}. One can see that the triple cluster amplitudes contribute about $+$3.8\% to the value of EFG for the ground state and $-$2.8\% for the excited state.

As is described above in the FS-CCSD calculation all 83 electrons were correlated. For the analysis we have also found that the contribution of sixty $1s..4f$ inner-core electrons of Bi contribute about 10\% and 7\% to the EFG values of the $6p^3$ $^4$S$^o_{3/2}$ and $6p^3$ $^2$P$^o_{3/2}$ states, respectively. This also justifies the procedure of taking into account triple cluster amplitudes, where these inner-core electrons were excluded from the electronic correlation calculation.

As one can see from Table~\ref{TRes} there is very good agreement between the values of the quadrupole moment of $^{209}$Bi extracted from hyperfine constants for the two considered electronic states. One can also stress the role of triple cluster amplitudes: they contribute in an opposite way to the EFG values of the two considered electronic states. In other words, the values of the quadrupole moments with EFG values obtained within the FS-CCSD approach differ more significantly than in the case when the triple amplitudes are included. This is not surprising and can be explained by the fact that the inclusion of (full) triple excitations is required to describe accurately a system with three valence 6p electrons.

Table~\ref{TRes} gives the effect of the Breit interaction. For the ground state, it contributes about 2\% to the total value of EFG, while for the excited state, its contribution is smaller -- 0.5\%. These contributions are strongly dominated by the Gaunt part, while the retardation part can be neglected. We have also estimated  contribution of quantum electrodynamics effects using the approach, developed in Ref.~\cite{Skripnikov:2021a} and found that it can be neglected for both states.

To estimate the uncertainty of the calculated EFG values we assume that the uncertainty of the basis set correction is smaller than 50\% of its value and the contribution of correlation effects beyond the FS-CCSDT model can hardly be notably larger than 50\% of the contribution of triple amplitudes~\cite{Skripnikov:17a,Skripnikov:16b,Skripnikov:15c,Skripnikov:17c}. For the ground and excited states this leads to the uncertainties of 9 mb and 6 mb for $Q(^{209}$Bi), respectively. As it can be seen from Table~\ref{TRes} the experimental uncertainties of the $B$ constants are negligible. For the $^{209}$Bi we average the value of $Q$ over two considered electronic states and obtain the final value of $Q(^{209}{\rm Bi})=-418(6)$ mb, employing the uncertainty obtained for the second state.

Our final value of $Q(^{209}$Bi) coincides within the uncertainty with the molecular results~\cite{Quevedo:2013,Avijit:2016}: $Q=-420(8)$~mb~\cite{Quevedo:2013} and $Q=-415$~mb~\cite{Avijit:2016}. It is also in the reasonable agreement with the independently measured values of $Q$ for muonic and pionic atoms: -370(26)~mb~\cite{Lee:1972,Raghavan:89}, -500(80)~mb~\cite{Beetz:1978}, but our uncertainty is much lower.

In Ref.~\cite{Pearson:2000} the hyperfine constants $B$ have been obtained and compiled from previous studies~\cite{Campbell:95,Alpert:62} for a large number of shot-lived isotopes of Bi in the ground electronic states $^4$S$^o_{3/2}$ (see also Ref.~\cite{Schmidt:2018} for an accurate measurement of $B(^{208}$Bi). Some of these constants were measured directly and others were rescaled, see Ref.~\cite{Pearson:2000} for details. Using the calculated value of EFG for the ground state of Bi it is possible to deduce the updated values of the electric quadrupole moments of these isotopes. Results are given in Table~\ref{TRes2}.

\begin{table}[]
\caption{The experimental values of electric quadrupole hyperfine constants $B$ for ground electronic states $^4$S$^o_{3/2}$ of short-lived Bi isotopes with mass numbers A and deduced values of the quadrupole moments of these isotopes. The experimental data for $^{208}$Bi isotope is taken from Ref.~\cite{Schmidt:2018}, while for others from Ref.~\cite{Pearson:2000}. The first value in parentheses for $Q$ corresponds to the uncertainty of the $B$ constant, the second value corresponds to the uncertainty of the theoretical value of EFG. In the case of $^{202}$Bi, results for two possible spin assignments are shown.}
\begin{tabular*}{0.45\textwidth}%
{@{\extracolsep{\fill}}llll}
\hline
\hline
A    & $I^{\pi}$ & $B$, MHz & $Q$, mb \\
\hline
202 & (5$^+$)      & -592(48)    &  -813(66)(17)   \\
202     & (6$^+$)      & -718(50)    &  -987(69)(21)   \\
203 & 9/2$^-$   & -549(40)    &  -754(55)(16)   \\
204 & 6$^+$     & -400(120)   &  -550(165)(12)   \\
205     & 9/2$^-$   & -481(10)    &  -661(14)(14)   \\
206     & 6$^+$     & -318(20)    &  -437(27)(9)   \\
207     & 9/2$^-$   & -449(27)    &  -616(37)(13)   \\
208     & 5$^+$     & --358.6(3.9)  &  -493(6)(11)    \\
210     & (1$^-$)   & 112.38(3)      &  154(0.04)(3)   \\
210$^m$ & 9$^-$     & -387(40)    &  -532(55)(11)   \\
212     & (1$^-$)   & 80(225)     &  110(309)(2)   \\
213     & 9/2$^-$   & -491(25)    &  -675(34)(14)   \\
\hline
\end{tabular*}
\label{TRes2}
\end{table}

\section{Conclusion}
The present atomic calculation performed at very high level of theory, the FS-CCSDT method within the Dirac-Coulomb-Breit Hamiltonian, leads to the value of $Q(^{209}$Bi), which is very close to the previous molecular studies. The experimental values for two different atomic electronic states were studied to get the value of $Q(^{209}$Bi). It is shown that the inclusion of triple cluster amplitudes is of crucial importance to get almost the same values of the extracted NQM's for both atomic levels. Obtained theoretical data allowed us also to refine the values of NQM of unstable $^{202-213}$Bi isotopes.

\begin{acknowledgments}
    Electronic structure calculations have been carried out using computing resources of the federal collective usage center Complex for Simulation and Data Processing for Mega-science Facilities at National Research Centre ``Kurchatov Institute'', http://ckp.nrcki.ru/.
    
    Calculations of the EFG values within the Dirac-Coulomb Hamiltonian were supported by the Russian Science Foundation (Grant No. 19-72-10019). Calculations of the Breit interaction contribution were supported by Russian Foundation for Basic Research (Grant No. 20-32-70177). Solution of Dirac-Fock equations was supported by the foundation for the advancement of theoretical physics and mathematics ``BASIS'' grant according to Projects No. 20-1-5-76-1 and 18-1-3-55-1.  The development of the version of FS-RCC code ({\sc exp-t}) accounting for the triple excitations is supported by the Russian Science Foundation (Grant No. 20-13-00225). The work to predict quadrupole moments of short-lived isotopes was supported by RFBR according to the research project No. 19-02-00005.
\end{acknowledgments}


\end{document}